# Significant improvement in sensitivity of an anomalous Nernst heat flux sensor by composite structure

Hiroto Imaeda(今枝寛人)[1], Reiji Toida(樋田怜史)[1], Tsunehiro Takeuchi(竹内恒博)[1], Hiroyuki Awano(粟野博之)[1], and Kenji Tanabe(田辺賢士)[1,a]

[1]Toyota Technological Institute, Nagoya, 468-8511, Japan

[a]Author to whom correspondence should be addressed: tanabe@toyota-ti.ac.jp

**Abstract (184 words)**

Heat flux sensors (HFS) have attracted significant interest for their potential in managing waste heat efficiently. A recently proposed HFS, that works on the basis of the anomalous Nernst effect (ANE), offers several advantages in its simple structure leading to easy fabrication, low cost, and reduced thermal resistance. However, enhancing sensitivity through traditional material selection is now challenging due to a small number of materials satisfying the required coexistence of a large transverse Seebeck coefficient and low thermal conductivity. In this study, by utilizing composite structures and optimizing the device geometry, we have achieved a substantial improvement in the sensitivity of an ANE-based HFS. We developed composite structures comprised of a plastic substrate with an uneven surface and three-dimensional (3D) uneven TbCo films, fabricated using nanoimprint techniques and sputtering. This approach resulted in a sensitivity that is approximately four times greater than that observed in previous studies. Importantly, this method is independent of the material properties and can significantly enhance the sensitivity. Our findings could lead to the development of highly sensitive HFS devices and open new avenues for the fabrication of 3D devices.

Unlike an ordinary temperature sensor, a heat flux sensor (HFS) allows us to visualize heat flow by detecting heat flux density as a vector. This capability has garnered significant interest for its potential to thermal management applications leading to a low-energy-consuming sustainable society. Commercial HFS devices work with the Seebeck effect (SE), where an electric field is generated along the direction of applied heat flow. However, for enhancing the generated voltage, the device has to have an increased thickness along the heat-flow direction, which consequently increases the thermal resistance to significantly affect the heat flux. This is a notable disadvantage. The complex structure of SE-based devices consisting of many π-type p/n legs naturally results in a high fabrication cost to prevent the wide use of HFS devices.

To address these issues, Zhou *et al.* recently proposed an HFS design based on the anomalous Nernst effect (ANE)[1]. The ANE can be expressed as

$$\boldsymbol{E}_{\mathrm{ANE}} = S_{\mathrm{ANE}} \boldsymbol{\nabla} T \times \left(\frac{\boldsymbol{M}}{|\boldsymbol{M}|}\right), \qquad (1)$$

where $\boldsymbol{E}_{\mathrm{ANE}}$, $S_{\mathrm{ANE}}$, $\boldsymbol{\nabla} T$ and $\boldsymbol{M}$ are an electric field induced by the ANE, a transverse Seebeck coefficient, a temperature gradient and a magnetization vector, respectively[2-3]. The magnitude of $S_{\mathrm{ANE}}$ is related to Berry's curvature in band structure. The key advantage of ANE-based HFSs is that the ANE electric field is induced perpendicular to the direction of heat flow. This orientation allows the device to maintain a thinner profile in the heat-flow direction, effectively reducing thermal resistance, and producing a larger voltage simply by a longer length in the direction perpendicular both to $\boldsymbol{\nabla} T$ and $\boldsymbol{M}$. Furthermore, ANE-based HFS devices consisting of a patterned single layer of ANE material are easier to fabricate due to their reliance on thin film technologies.

$S_{\mathrm{ANE}}$ is one of the most important parameters for ANE-based HFS devices. Historically, $S_{\mathrm{ANE}}$ was understood to be proportional to saturation magnetization[3]. However, a groundbreaking study by Ikhlas *et al.* uncovered a relatively large $S_{\mathrm{ANE}}$ of 0.6 μVK$^{-1}$ at room temperature in antiferromagnetic Mn$_3$Sn, despite its ultralow saturation magnetization[4]. This notable ANE is attributed to enhanced Berry curvature at Weyl points near the Fermi energy, a discovery that has catalyzed numerous subsequent research efforts[5-20]. Particularly significant is the finding that the full-Heusler ferromagnet Co$_2$MnGa exhibits a large $S_{\mathrm{ANE}}$ of 6 μVK$^{-1}$, setting a current world record at room temperature. He *et al.* also reported a magnon-induced ANE in MnBi, which differs from the Berry curvature-induced ANE[15]. These developments underscore a deepening understanding and expanding technological potential of ANE across diverse materials.

An ANE-based HFS exhibits two distinct types of sensitivity: device sensitivity and material sensitivity[22]. Device sensitivity, defined within a specific fabricated device, is measured as the induced voltage per unit of heat flux density ($V_{\text{ANE}}/j_q$). The device sensitivity depends on the size of the sensor area and the device shape, and, therefore, it does not serve as a suitable parameter for evaluating materials. Material sensitivity, on the other hand, focuses solely on material parameters and is quantified as the induced electric field per unit of heat flux density ($E_{\text{ANE}}/j_q$), which remains consistent regardless of size. Thus, the evaluation of materials for HFS applications should be based primarily on $E_{\text{ANE}}/j_q$. Material sensitivity can be analytically derived using Fourier's law, $j_q = -\kappa \nabla T$, and the transverse Seebeck coefficient ($S_{\text{ANE}}$). The formula $E_{\text{ANE}}/j_q = S_{\text{ANE}}/\kappa$ integrates only material parameters: $S_{\text{ANE}}$ and the thermal conductivity $\kappa$. To achieve high material sensitivity, it is crucial that the material exhibits a low $\kappa$ together with a large $S_{\text{ANE}}$.

The demonstration of an ANE-based HFS has been performed by using materials with a large $S_{\text{ANE}}$ such as Fe-Al[1], Co$_2$MnGa[23], and Fe-Ga[24]. Although material sensitivity was not explicitly reported in these studies, we can be estimated to be approximately 0.20 μm/A for Fe$_{79}$Al$_{21}$, 0.22 μm/A for Co$_2$MnGa, and 0.16 μm/A for Fe$_{81}$Ga$_{19}$ from the reported device sensitivity and the shape of the devices. Conversely, some research groups have focused on materials with low thermal conductivity ($\kappa$) such as amorphous Sm-Co[25] and Gd-Co[22] alloys and material sensitivities were 0.18 μm/A for Sm$_{20}$Co$_{80}$ and -0.23 μm/A for Gd$_{24}$Co$_{76}$. Despite significant differences in $S_{\text{ANE}}$ and $\kappa$ among these materials, the ratio of $|S_{\text{ANE}}/\kappa|$ remains approximately 0.2 μm/A across all materials, indicating that the sensitivity improvement is limited in material exploration.

In this study, we present a novel approach to improve $E_{\text{ANE}}/j_q$, moving away from traditional material exploration. We have developed a composite material comprising a magnetic component and an ultralow-$\kappa$ polymer, structured into a land-groove form using the nanoimprint method. This innovative design significantly enhances sensitivity by directing heat flow into the magnetic material, achieving a sensitivity approximately four times greater than that of single materials previously reported. This method opens new avenues for fabricating three-dimensional(3D) devices with enhanced thermal sensing capabilities.

The conceptual framework of this study is depicted in Figs. 1(a-b). Figure 1(a) illustrates that when a uniform heat flux density $j_{q\_\text{in}}$ is applied to a single material from its top surface, the heat flows uniformly through the material, resulting in a heat flux density through the material ($j_q$) that equals $j_{q\_\text{in}}$. In contrast, Figure 1(b) shows that, in a composite material, heat flow is concentrated into the ANE material with higher thermal conductivity ($\kappa$), represented as $j_{q\_in} \ll j_q$. This design allows the heat flow to be focused into the magnetic material, which is surrounded by an ultralow-$\kappa$ polymer. This concentration of heat flow enhances the temperature gradient ($\nabla T$) in the ANE material, thereby enhancing the induced electric field per unit heat flux density ($E_{ANE}/j_{q\_in}$). Although the estimated value of $E_{ANE}/j_{q\_in}$ does not strictly represent the material sensitivity, but a pseudo-material sensitivity of the composite structure, as will be discussed later. To implement this design, we utilized a substrate with a 3D land-groove structure.

We fabricated this 3D land-groove structure on the surface of a plastic substrate (ZeonorFilm 16, produced by Nihon Zeon Corp.) using the nanoimprint method, as depicted in Fig. 1(c) and previously described in previous article[26]. Initially, a silicon mold featuring a fine land-groove structure was placed on the plastic substrate at room temperature. The mold and substrate were then heated to 453 K—above the glass transition temperature of the plastic substrate (436 K)—and maintained at this temperature for 10 minutes. Subsequently, they were pressed at 0.2 MPa for 4 minutes for transferring the precise land-groove structure onto the plastic substrate. After cooling, the substrate was separated from the mold and the obtained land-groove structure was exposed to ultraviolet light for 3 minutes for smoothing the surface, the finally obtained land-groove structure has widths of 5 μm, heights of 2, 5, and 10 μm, spaced 5 μm apart, fabricated over an area of 4×4 mm².

$SiN_x$ (10 nm)/$Tb_{10}Co_{90}$ ($d$)/$SiN_x$ (10 nm) films were deposited on the surface of the plastic substrate with the land-groove structure by an ultra-high vacuum magnetron sputtering system. The $Tb_{10}Co_{90}$ layer was deposited by a co-sputtering method. The base pressure and sputtering pressure during deposition are less than $3 \times 10^{-5}$ Pa and almost 0.2 Pa, respectively. Alloys like Tb-Co, which are amorphous ferrimagnetic materials that combine rare-earth and transition metals, are particularly valued for their ability to easily control various magnetic parameters, including the magnetization, angular momentum, magnetic anisotropy, easy magnetization axis, and domain size[27-32]. In this experiment, the Tb composition has been fixed at roughly 10% to clarify the effects of the structure rather than composition. It was verified via energy dispersive X-ray spectroscopy and composition deviation is about 3%. Film thickness varied from 5 to 200 nm; these values refer to films deposited on a flat substrate, noting that the actual thicknesses on the top and bottom surfaces and the sides of the 3D land-groove structure vary. The $SiN_x$ layers serve as protective barriers to prevent oxidation of the $Tb_{10}Co_{90}$ layer. Polymer resin (Zeocoat CP1010-14 made by Nihon Zeon Corp.) was coated with a spin coater on the surface of these samples. Its thickness is substantial relative to the height of the land-groove structure. Both the bottom plastic substrate and the top resin layer possess ultralow thermal conductivity ($\kappa$). Figure 1(c) provides a schematic cross-sectional view of the 3D land-groove structure. When heat is applied the out-of-plane direction, and concentrates on the high-aspect-ratio $Tb_{10}Co_{90}$ structures deposited on the side surfaces of the land-grooves.

Figure 1(d) shows cross-sectional transmission electron microscopy (TEM) images of one land structure, which was taken by the MST Corp. The $Tb_{10}Co_{90}$ thickness designed on a flat substrate $d$ is 100 nm, while actual measurements show thicknesses of approximately 85 nm on the top surface, 48 nm on the bottom, and 22 nm on the sides. The side thickness was nearly a quarter of the designed thickness.

The thermal conductivities ($\kappa$) of the plastic substrate and the top resin layer were experimentally determined, as shown in Table 1. For the plastic substrate, $\kappa$ was calculated from the thermal diffusivity ($\alpha$) measured by the cyclic heat method, specific heat ($C$) by differential scanning calorimetry (DSC), and density ($\rho$), giving $\kappa = \alpha C \rho$. The amorphous nature of the substrate implies isotropic thermal properties. The thermal conductivity $\kappa$ of the top resin layer was derived from the thermal effusivity ($b$) measured by time-domain thermoreflectance (Nano-TR, Netzsch Corp.), along with $C$ and $\rho$, resulting in $\kappa = b^2/C\rho$. The measured values of $\kappa$ for the plastic substrate and top resin layer are 0.18 W/mK and 0.47 W/mK, respectively. These values are significantly lower than those of typical metallic magnetic materials such as Co, Fe, and Tb-Co films, aligning with the properties expected of polymer materials.

The setup for measuring $E_{\text{ANE}}/j_{q\_\text{in}}$ is depicted in Fig. 2(a). The sample, placed between two Cu blocks working as heat reservoirs, is connected to a commercialized HFS and a heater. The magnetic field is adjusted while maintaining constant heat flow from the heater, detected by the HFS. ANE voltage measurements are conducted using Au electrodes deposited by a resistance-heating evaporator, electrical probes, and a nanovoltmeter. The magnetic field is oriented along the in-plane direction and perpendicular to the land-groove structure. For validation, $E_{\text{ANE}}/j_{q\_\text{in}}$ measurements on a Py film deposited on a flat plastic substrate yielded +0.018 μm/A, aligning with the ratio of the transverse Seebeck coefficient to thermal conductivity in previous findings[33-34].

Figure 2(b, c) displays the variation in $V_{\text{ANE}}$ with magnetic field for $Tb_{10}Co_{90}$ films ($d = 20$ nm) deposited on both flat and land-groove structured substrates (5 μm in height). On the flat substrate, $V_{\text{ANE}}$ exhibits a sign reversal near zero magnetic field due to the in-plane orientation of magnetization in Fig. 2(b). A similar sign reversal is noted for the land-groove structure, caused by the in-plane orientation of magnetization

at the top and bottom surfaces of the land-groove structure. As the magnetic field increases, $V_{\text{ANE}}$ grows and ultimately saturates. This pattern was obtained because the magnetization of the TbCo film on the side surfaces, initially in-plane at zero magnetic field, aligns with the magnetic field direction as it increases, illustrated in Fig. 2(e). Figure 2(f) explores the relationship between heat flux density and $V_{\text{ANE}}$ at a magnetic field of 10 kOe, showing that $V_{\text{ANE}}$ is directly proportional to heat flux density, with the slope related to the pseudo-material sensitivity.

We investigated the thickness dependence of the sensitivity $E_{\text{ANE}}/j_{q\_\text{in}}$ of the $Tb_{10}Co_{90}$ films on both flat substrates and substrates with 10 μm land-groove height as shown in Fig. 3(a). On flat substrates, $E_{\text{ANE}}/j_{q\_\text{in}}$ remains nearly constant at about 0.097 μm/A, showing only minor variations presumably due to slight compositional differences in $Tb_{10}Co_{90}$. However, for substrates with a 10 μm height, $E_{\text{ANE}}/j_{q\_\text{in}}$ significantly varies with film thickness. For films thicker than 40 nm, $E_{\text{ANE}}/j_{q\_\text{in}}$ decreases with increasing thickness. This trend is attributed to a decrease in thermal resistance of the $Tb_{10}Co_{90}$ deposited on the side surfaces, which in turn reduces the thermal resistance of the whole sample and lowers the temperature gradient. When the designed thickness is smaller than 40 nm, $E_{\text{ANE}}/j_{q\_\text{in}}$ paradoxically decreases with reducing thickness—an unexpected result that may stem from surface roughness. As shown in the magnified view at the land-groove side in Fig. 1(d), the $Tb_{10}Co_{90}$ film on the side exhibits significant roughness, potentially leading to reduced thermal conductivity. Previous studies reported that the thermal conductivity in $Tb_{21}Fe_{73}Co_6$ alloys is approximately 5 W/mK[35], which is nearly ten times higher than that of the plastic substrate and polymer layer. However, if the thermal conductivity of $Tb_{10}Co_{90}$ on the sides diminishes to levels closer to those of the substrate and polymer, heat flow concentration is hindered. Consequently, thinner films, being more influenced by surface effects, exhibit lower $E_{\text{ANE}}/j_{q\_\text{in}}$. In addition, the films may become discontinuous layers by decreasing thickness. In this case too, $E_{\text{ANE}}/j_{q\_\text{in}}$ decreases like the reduction in thermal conductivity owing to suppression of thermal conduction of the $Tb_{10}Co_{90}$ films. Notably, $E_{\text{ANE}}/j_{q\_\text{in}}$ peaks at 0.80 μm/A when the film thickness is exactly 40 nm.

We investigated the impact of land-groove structure height on the sensitivity $E_{\text{ANE}}/j_{q\_\text{in}}$, as shown in Fig. 3(b). Our findings indicate that sensitivity increases with the height of the land-groove structure, which can be attributed to two main factors. Firstly, at greater heights, heat flow tends to be channeled through the $Tb_{10}Co_{90}$ film, enhancing the temperature gradient ($\nabla T$) in the $Tb_{10}Co_{90}$ film. This is because the elevated structure promotes the shunting of heat through the $Tb_{10}Co_{90}$ rather than directly through the polymer, thereby concentrating the heat flow within the metallic layer and increasing its effect on sensitivity. Secondly, the suppression of the thickness of the $Tb_{10}Co_{90}$ film on the side surfaces contributes to increased sensitivity. As the height of the land-groove structure increases, the surface area of this structure also increases, leading to a reduction in the actual thickness of the $Tb_{10}Co_{90}$ film, despite the same designed thickness. This reduction in thickness, compounded by the roughness of the films, diminishes the thermal conductivity, further enhancing $\nabla T$ in the $Tb_{10}Co_{90}$ film deposited on the side surfaces. These factors lead not only to the larger heat flux passing through the $Tb_{10}Co_{90}$ film but also to the reduction in its ability to conduct heat, thereby ultimately boosting the sensitivity of the device.

Let us now compare our results obtained in the uneven structure with the previous findings. The sensitivity measured in this study differs significantly from the typical material sensitivity because the heat flux density ($j_{q\_in}$) is not solely determined by the $Tb_{10}Co_{90}$ film but by the entire device structure. Nonetheless, if a device incorporates such a composite structure, the high sensitivity observed in our experiment is as valuable as traditional material sensitivity.

Figure 4(a) provides a schematic illustration of the device structure, where blue and yellow wires represent ANE materials with opposite signs of $S_{ANE}$. When these wires are flat, the device sensitivity can be directly inferred from the material sensitivity. Conversely, when the wires have uneven structures, the device sensitivity aligns with the values obtained in our experiments. The material sensitivities reported previously are plotted in Fig. 4 (b) together with our current results. In sharp contrast to the previously reported material sensitivity staying around 0.2 μm/A, our composite material exhibited a superior sensitivity of $|E_{ANE}/j_{q\_in}|$ = 0.80 μm/A.

Importantly, the device design employed in this study is not confined to the use of $Tb_{10}Co_{90}$ alloys; it can be applied to any other materials. Previous research has emphasized the necessity of combining a large transverse Seebeck coefficient with low thermal conductivity to achieve highly sensitive HFS[22]. However, our findings suggest that low thermal conductivity is not the necessity for realizing a high sensitivity due to the overall suppression of thermal conductivity by the composite structure. Rather than focusing on low thermal conductivity, the critical factor now becomes the fabrication technique of ferromagnetic films that exhibit a large transverse Seebeck coefficient and magnetic anisotropy perpendicular to the side of the land-groove structure.

In conclusion, we have significantly enhanced the sensitivity of ANE-based HFS by employing a novel approach centered on optimizing geometry, rather than relying on conventional material searches. Utilizing the nanoimprint method, we created a 3D land-groove structure composed of a $Tb_{10}Co_{90}$ film coupled with an ultralow-κ polymer. Notably, when the $Tb_{10}Co_{90}$ film was adjusted to a thickness of 40 nm, the sensitivity ($E_{ANE}/j_{q\_in}$) achieved in the land-groove structure was 0.80 μm/A—nearly four times greater than that observed with single materials previously reported. This innovative method not only departs from traditional techniques but also paves the way for the development of 3D devices with superior heat flux sensing capabilities.

This work was partly supported by the Paloma Environmental Technology Development Foundation.

## AUTHOR DECLARATIONS
Conflict of Interest

The authors have no conflicts to disclose.

Author Contributions

**Hiroto Imaeda**: Data Curation (lead); Investigation (lead); Formal Analysis (lead); Writing/Original Draft Preparation (equal); Writing/Review & Editing (supporting); **Reiji Toida**: Data Curation (supporting); Formal Analysis (supporting); Methodology (supporting); **Tsunehiro Takeuchi**: Investigation (supporting); Resources (equal); Writing/Review & Editing (supporting); **Hiroyuki Awano**: Investigation (supporting); Resources (equal); Writing/Review & Editing (supporting); **Kenji Tanabe**: Conceptualization (lead); Data Curation (supporting); Formal Analysis (supporting); Funding Acquisition (lead); Investigation (supporting); Methodology (lead); Resources (equal); Writing/Original Draft Preparation (equal); Writing/Review & Editing (lead)

## DATA AVAILABILITY

The data that support the findings of this study are available from the corresponding author upon reasonable request.

## REFERENCES

1) W. Zhou and Y. Sakuraba, Applied Physics Express **13**, 043001 (2020).
2) Y. Sakuraba, Scripta Materialia **111**, 29–32 (2016).
3) M. Mizuguchi and S. Nakatsuji, Science and Technology of Advanced Materials, **20**, 262, (2019).

4) M. Ikhlas, T. Tomita, T. Koretsune, M.-T. Suzuki, D. Nishio-Hamane, R. Arita, Y. Otani and S. Nakatsuji, Nature Physics **13**, 1085-1090 (2017).

5) A. Sakai, Y. P. Mizuta, A. A. Nugroho, R. Sihombing, T. Koretsune, M.-T. Suzuki, N. Takemori, R. Ishii, D. Nishio-Hamane, R. Arita, P. Goswami and S. Nakatsuji, Nature Physics **14**, 1119-1124 (2018).

6) S. N. Guin, K. Manna, J. Noky, S. J. Watzman, C. Fu, N. Kumar, W. Schnelle, C. Shekhar, Y. Sun, J. Gooth, and C. Felser, NPG Asia Materials **11**, 16 (2019).

7) K. Sumida, Y. Sakuraba, K. Masuda, T. Kono, M. Kakoki, K. Goto, W. Zhou, K. Miyamoto, Y. Miura, T. Okuda, and A. Kimura, Communications Materials **1**, 89 (2020).

8) T. Asaba, V. Ivanov, S. M. Thomas, S. Y. Savrasov, J. D. Thompson, E. D. Bauer, and F. Ronning, Science Advance **7**, eabf1467 (2021).

9) H. Nakayama, K. Masuda, J. Wang, A. Miura, K. Uchida, M. Murata, and Y. Sakuraba, Physical Review Materials **3**, 114412 (2019).

10) A. Sakai, S. Minami, T. Koretsune, T. Chen, T. Higo, Y. Wang, T. Nomoto, M. Hirayama, S. Miwa, D. Nishio-Hamane, F. Ishii, R. Arita and S. Nakatsuji, Nature **581**, 53 (2020).

11) Y. Sakuraba, K. Hyodo, A. Sakuma, and S. Mitani, Phys. Rev. B **101**, 134407 (2020).

12) H. Yang, W. You, J. Wang, J. Huang, C. Xi, X. Xu, C. Cao, M. Tian, Z.-A. Xu, J. Dai, and Y. Li, Phys. Rev. Materials **4**, 024202 (2020).

13) T. Chen, T. Tomita, S. Minami, M. Fu, T. Koretsune, M. Kitatani, I. Muhammad, D. Nishio-Hamane, R. Ishii, F. Ishii, R. Arita, and S. Nakatsuji, Nat. Commun. **12**, 572 (2021).

14) D. Khadka, T. R. Thapaliya, S. H. Parra, J. Wen, R. Need, J. M. Kikkawa, and S. X. Huang, Phys. Rev. Materials **4**, 084203 (2020).

15) B. He, C. Şahin, S. R. Boona, B. C. Sales, Y. Pan, C. Felser, M. E. Flatté, J. P. Heremans, Joule **5**, 3057-3067 (2021).

16) T. Chen, S. Minami, A. Sakai, Y. Wang, Z. Feng, T. Nomoto, M. Hirayama, R. Ishii, T. Koretsune, R. Arita, S. Nakatsuji, Science Advance **8**, eabk1480 (2022).

17) K. Fujiwara, Y. Kato, H. Abe, S. Noguchi, J. Shiogai, Y. Niwa, H. Kumigashira,

Y. Motome, and A. Tsukazaki, Nature Communications **14**, 3399 (2023).

18) S. Roychowdhury, A. M. Ochs, S. N. Guin, K. Samanta, J. Noky, C. Shekhar, M. G. Vergniory, J. E. Goldberger, and C. Felser, Adv. Mater. **34**, 2201350 (2022).

19) Y. Pan, C. Le, B. He, S. J. Watzman, M. Yao, J. Gooth, J. P. Heremans, Y. Sun and C. Felser, Nature Materials **21**, 203 (2022).

20) M. Odagiri, H. Imaeda, A. Yagmur, Y. Kurokawa, S. Sumi, H. Awano, and K. Tanabe, Applied Physics Letters **124**, 142403 (2024).

21) S. Noguchi, K. Fujiwara, Y. Yanagi, M.-T. Suzuki, T. Hirai, T. Seki, K.-i. Uchida and A. Tsukazaki, Nature Physics **20**, 254–260 (2024).

22) M. Odagiri, H. Imaeda, A. Yagmur, Y. Kurokawa, S. Sumi, H. Awano, and K. Tanabe, arXiv:2402.04259.

23) K.-i. Uchida, W. Zhou, and Y. Sakuraba, Appl. Phys. Lett. **118**, 140504 (2021).

24) H. Tanaka, T. Higo, R. Uesugi, K. Yamagata, Y. Nakanishi, H. Machinaga, S. Nakatsuji, Advanced Materials, **35**, 2303416 (2023).

25) R. Modak, Y. Sakuraba, T. Hirai, T. Yagi, H. S.-Amin, W. Zhou, H. Masuda, T. Seki, K. Takanashi, T. Ohkubo and K.-i. Uchida, Sci. Technol. Adv. Mater. **23**, 777 (2022).

26) T. Asari, R. Shibata, and H. Awano, AIP Advances 7, 055930 (2017).

27) C. D. Stanciu, A. V. Kimel, F. Hansteen, A. Tsukamoto, A. Itoh, A. Kirilyuk, and T. Rasing, Phys. Rev. B **73**, 220402(R) (2006).

28) K.-J. Kim, S. K. Kim, Y. Hirata, S.-H. Oh, T. Tono, D.-H. Kim, T. Okuno, Woo S. Ham, S. Kim, G. Go, Y. Tserkovnyak, A. Tsukamoto, T. Moriyama, K.-J. Lee and T. Ono, Nature Materials **16**, 1187–1192 (2017).

29) A. Yagmur, S. Sumi, H. Awano, and K. Tanabe, Physical Review Applied **117**, 242407 (2020).

30) S. Fukuda, H. Awano, and K. Tanabe, Applied Physics Letters **116**, 102402 (2020).

31) A. Yagmur, S. Sumi, H. Awano, and K. Tanabe, Physical Review B **103**, 214408 (2021).

32) S. Kuno, S. Deguchi, S. Sumi, H. Awano, and K. Tanabe, APL Machine Learning **1**, 046111 (2023).

33) T. Yamazaki, T. Seki, R. Modak, K. Nakagawara, T. Hirai, K. Ito, K.-i. Uchida, and K. Takanashi, Phys. Rev. B **105**, 214416 (2022).

34) T. Kikkawa, K. Uchida, S. Daimon, Y. Shiomi, H. Adachi, Z. Qiu, D. Hou, X.-F. Jin, S. Maekawa, and E. Saitoh, Phys. Rev. B **88**, 214403 (2013).

35) P. E. Hopkins, M. Ding and J. Poon, J. Appl. Phys. **111**, 103533 (2012).

**Figure and Table Caption**

Fig. 1(a-b) Schematic illustrations of heat flows. In (a), for a single material, the heat flux density $j$ through the magnetic material is equal to the input heat flux density $j_{in}$. In (b), for a composite material, $j$ through the magnetic material is significantly enhanced compared to $j_{in}$. Blue areas represent magnetic material, and gray areas indicate material with ultralow thermal conductivity. (c) Step-by-step schematic diagram of the sample fabrication process: starting with a Si mold with an uneven structure, pressing it onto a plastic substrate above its glass transition temperature, transferring the mold structure to the substrate, and finishing with the deposition of a magnetic film on the substrate surface. (d) Cross-sectional TEM images within a land-groove structure, highlighting the TbCo films in black.

Table 1 Thermal characteristics of the substrate and resin layer.

Fig. 2(a) Schematic of the measurement setup. (b-c) The ANE voltage as a function of a magnetic field for a flat sample(b) and unevenness sample(c) with 10 μm height at room temperature, with a heat flux density of approximately 26 kW/m². (d-e) Illustrations of the relationship among heat flow, magnetic field and magnetization. (f) The ANE voltage as a function of heat flux density for unevenness sample with 10 μm height at room temperature.

Fig. 3(a) Sensitivity as a function of designed thickness, where square dots are results from flat substrates and round dots are from 10 μm land-groove height substrates. The red curve is a guide to the eye. (b) Sensitivity as a function of land-groove height.

Fig. 4(a) Schematic of a heat flux device, showing blue and yellow wires as magnetic materials with opposite $S_{ANE}$ signs. In the composite structure, these wires have uneven structure. (b) Summery graph of $|E_{ANE}/j_{in}|$. Sensitivity data for $Fe_{79}Ga_{21}$[24], $Sm_{20}Co_{80}$[25], $Fe_{81}Al_{19}$[1], $Co_2MnGa$[23], and $Gd_{24}Co_{76}$[22] are referenced from respective publications. "Unevenness" refers to data obtained in this work.

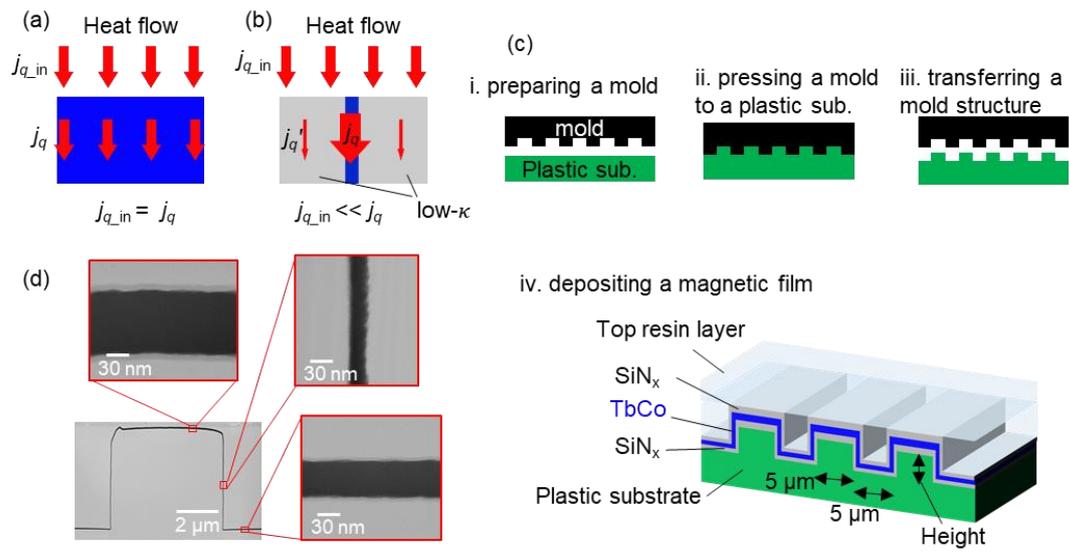

Figure 1

|  | Thermal diffusivity $m^2s^{-1}$ | Thermal effusivity $Js^{1/2}m^2K$ | Specific heat $Jg^{-1}K^{-1}$ | Density $gcm^{-3}$ | Thermal conductivity $Wm^{-1}K^{-1}$ |
|---|---|---|---|---|---|
| Plastic substrate | $1.85 \times 10^{-7}$ |  | 1.09 | 0.914 | 0.18 |
| Top resin layer |  | 553 | 0.88 | 0.743 | 0.47 |

Table 1

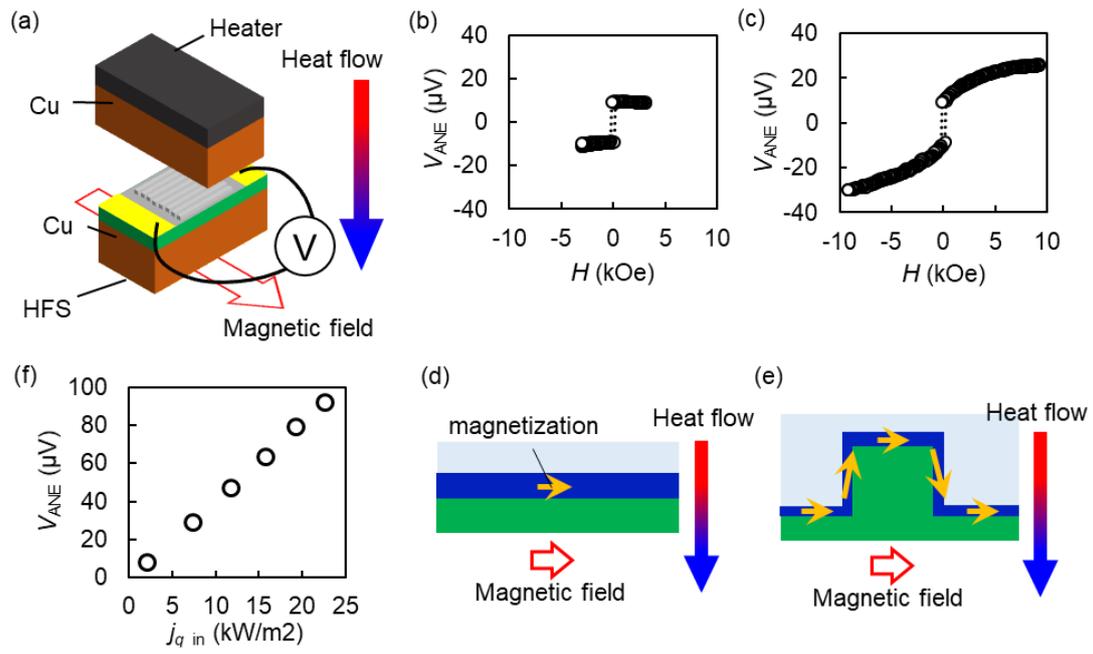

Figure 2

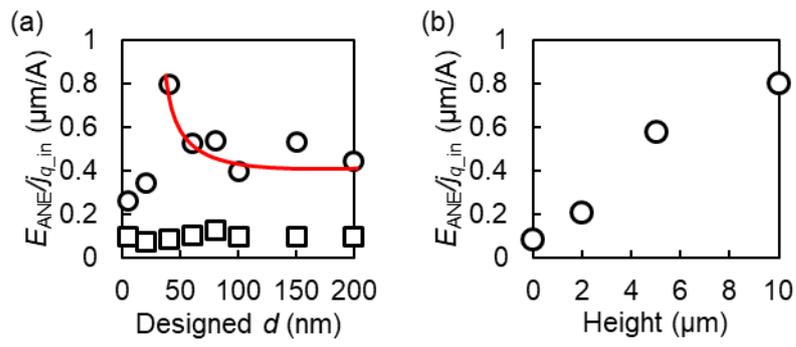

Figure 3

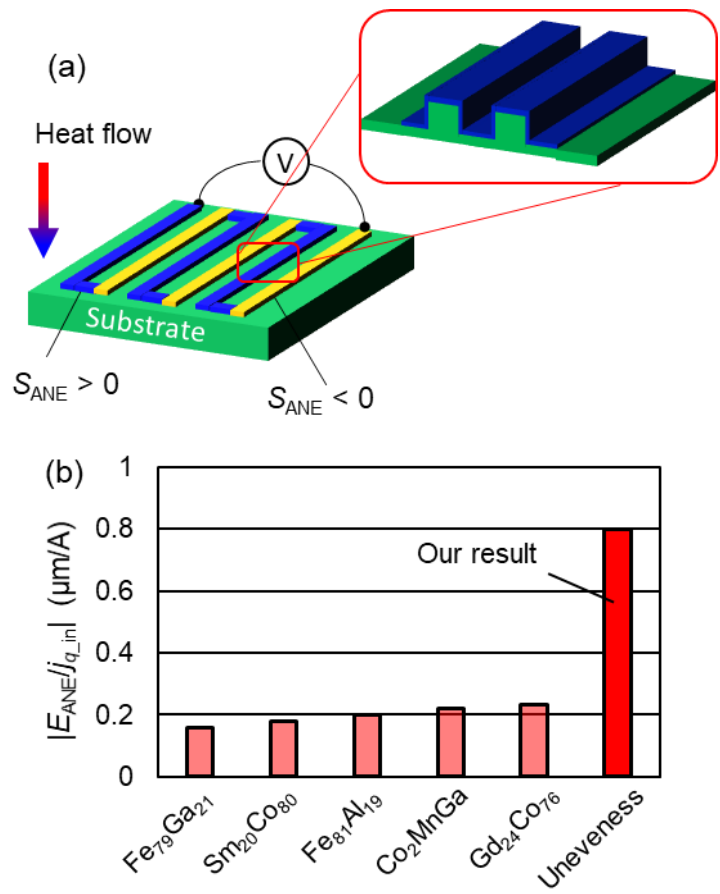

Figure 4